\documentclass[]{jfm}

\usepackage{graphicx}
\usepackage{newtxtext}
\usepackage{newtxmath}
\usepackage{natbib}
\usepackage{hyperref}
\hypersetup{
    colorlinks = true,
    urlcolor   = blue,
    citecolor  = black}

\newcommand{\RomanNumeralCaps}[1]
\linenumbers

\newcommand\bu{\boldsymbol{u}}
\newcommand\be{\boldsymbol{e}}
\newcommand{\pd} [2] {\frac{\partial #1}{\partial #2}}

\title{Scaling in two-dimensional Rayleigh-B\'enard convection}

\author{ Erik Lindborg
  \corresp{\email{erikl@mech.kth.se}}  }

\affiliation{Department of Engineering Mechanics, KTH, Osquars backe 18, SE-100 44, Stockholm, Sweden}

\begin{document}
\maketitle
\begin{abstract}
An equation for the evolution of mean kinetic energy, $ E $,  in a 2-D or 3-D Rayleigh-B\'enard system with domain height $ L $ is derived. 
 Assuming classical Nusselt number scaling, $ Nu \sim Ra^{1/3} $, and that mean enstrophy, in the absence of a downscale energy cascade,  scales as $\sim E/L^2 $, we
find that the Reynolds number scales as $ Re  \sim Pr^{-1}Ra^{2/3} $ in the 2-D system, where $ Ra $ is the Rayleigh number and $ Pr $ the Prandtl number, which is a much stronger scaling than in the 3-D system. 
Using the evolution equation and the Reynolds number scaling, it is shown that 
$ \tilde{\tau}  >  c Pr^{-1/2}Ra^{1/2} $, where 
 $ \tilde{\tau} $ is the non-dimensional convergence time scale and $ c $ is a non-dimensional constant. 
 For  the 3-D system, we make the estimate $ \tilde{\tau}  \gtrsim  Ra^{1/6} $ for $ Pr = 1 $. It is estimated that the total computational cost of reaching the high $ Ra $ limit in a simulation is comparable between 2-D and 3-D. 
 The results of the analysis are compared to DNS data and it is concluded that the theory of the `ultimate state'  is not valid in 2-D.  
 Despite the big difference between the 2-D and 3-D systems in the scaling of $ Re $ and $ \tilde{\tau} $, the Nusselt number scaling is similar. This observation supports the hypothesis of \citet{Malkus54} that the heat transfer is not regulated by the dynamics in the interior of the convection cell, but by the dynamics in the boundary layers. 
 
 \end{abstract}

\section{Introduction}
The problem of scaling in Rayleigh-B\'enard convection (RBC) has a long history represented by a huge body of literature. In this introduction, we will not make any attempt to review this literature but concentrate on an issue which is of great relevance for the current debate, namely the differences and similarities between the 2-D and 3-D Rayleigh-B\'enard systems.  For general reviews on the problem the reader is referred to  \citet{Siggia94} and \citet{Chilla12}, and for a review on the current debate the reader is referred to \citet{Lindborg23}. The essay by \citet{Doering20} is also highly recommended. 

Despite the fundamental difference between two- and three-dimensional turbulence, a clear distinction is rarely made between 2-D and 3-D in the theoretical discussion of RBC. The arguments for and against classical Nusselt-number scaling, $ Nu \sim Ra^{1/3} $ \citep{Malkus54},  and non-classical scaling, $ Nu \sim Ra^{1/2} $  \citep{Spiegel63}, $ Nu \sim Ra^{1/2} [ \ln (Ra)]^ {-3/2} $ \citep{Kraichnan62} or  $ Nu \sim Ra^{2/7} $ \citep{Castaing89}, are most often discussed as if they were equally relevant for the 2-D and 3-D systems. Making a series of 2-D DNS, \citet{Zhu18} claim that they found evidence of a transition
to the so called `ultimate state' of heat transfer predicted by \citet{Kraichnan62}. \citet{Doering19}  question the curve fitting of \cite{Zhu18} and conclude that their data are consistent with classical Nusselt-number scaling in accordance with the theory of \citet{Malkus54}. In a reply, Zhu et al. (2019) answer that they have carried out two more simulations at even higher Ra and that their curve fit now shows ‘overwhelming evidence’ of a transition to the ultimate state and that they have `irrefutably settled the issue'. These claims were repeated by \citet{Lohse24}.  Another way of questioning such claims is to point out that the theory of the ultimate state is based on assumptions that may be valid in three dimensions, but, most likely, are not valid in two dimensions. In particular, the theory is based on the assumption that the boundary layers in a convection cell in the limit of high $ Ra $ are of the same type as classical turbulent boundary layers in a 3-D shear flow and that the friction law of such a boundary layer can be approximated as
\begin{equation} \label{Friction} 
u_{\tau} = {\overline{\kappa}} {U} [\ln(Re)]^{-1} \, , 
\end{equation} 
where $ u_\tau $ is the friction velocity, $ U $ is the free stream velocity and  $ \overline{\kappa} $ is the K\'arm\'an constant. The modified theory of  \citet{Grossmann11} is based on the same general assumption but uses 
a slightly modified form of (\ref{Friction}).  In order to apply the theory to the 2-D case it must be assumed that a friction law of this type is valid also in 2-D.  However, there is no reason to believe this, because 2-D and 3-D turbulence are fundamentally different. 
\citet{Falko18} investigated 2-D Couette and Poiseuille flow and presented strong numerical and theoretical evidence showing that 2-D Couette flow will never become turbulent no matter how large the Reynolds number is,  while the high Reynolds-number  Poiseuille flow indeed is turbulent but exhibits a completely  different boundary layer structure than the corresponding 3-D flow, with a friction law of the form $ u_{\tau} \sim U Re^{-1/4} $. This friction law was also found in 2-D turbulence experiments in soap films by  \citet{Tran10}.   \cite{Zhu18} claim that their simulations, indeed,  show that the boundary layers in 2-D RBC are of the same form as in a 3-D shear flow, including a logarithmic dependence  of an appropriately defined mean velocity, $ U$. However, instead of the classical logarithmic law, they claim that they have discovered a law of the form 
\begin{equation} \label{LOG}
U(z) = u_{\tau} \left ( \frac{1}{\overline{\kappa}}  \ln \left (\frac{u_{\tau} z}{\nu} \right ) + B(Re) \right) \, , 
\end{equation} 
where $ z $ is the distance from the wall and $ B(Re) $ is an increasing function of $ Re $, instead of being a constant as in a standard 3-D boundary layer. It is clear that such a Reynolds number dependence is not consistent  with (\ref{Friction}) or the corresponding friction law used by \cite{Grossmann11} which both are derived under the assumption that $ B $ is a constant. 
This is not discussed by \citet{Zhu18}. 

The ultimate state theory is developed by using (\ref{Friction}) and some additional assumptions to derive a closed set of equations relating the Nusselt, Reynolds and Rayleigh numbers, where the 
Reynolds number is based on the characteristic velocity fluctuations in the interior of the convection cell. 
The scaling relation between $ Re $ and $ Ra $ reads:  $ Re \sim Ra^{1/2} $, including some Prandtl number dependent factor and possibly some logarithmic factor, which are different between different versions of the theory \citep[e.g.][]{Kraichnan62, Grossmann11, Shishkina24}. The prediction for the Reynolds number  is central. 
 If it can be shown that it is not valid, the theory can be regarded as falsified. 

 It is often observed in numerical simulations  that $ Re $ has a stronger scaling with $ Ra $ in 2-D as compared to 3-D, with $ Re \sim Ra^{\beta} $, where $ \beta > 1/2 $ instead of $ \beta < 1/2 $ as most often is observed in 3-D.
 \citet{Poel13} note  this but do not analyse the reason for the difference or discuss if it will prevail in the limit of high $ Ra $. 
 Exact steady but unstable solutions to the 2-D problem have been analysed by \citet{Chini09, Wen20}  for stress-free boundary conditions and by \citet{Waleffe15, Wen22} for no-slip boundary conditions. 
 The solutions show classical Nusselt number scaling, while the Reynolds number scaling is quite different between the two cases. For stress-free boundary conditions a clean scaling  of the form $ Re \sim Pr^{-1} Ra^{2/3} $ is obtained. 
 It should be pointed out that for a 2-D system with stress-free boundary conditions \citet{Whitehead11} rigorously proved that the Nusselt number is limited by $ 0.2891 Ra^{5/12} $, ruling out $ Ra^{1/2} $-scaling, including possible logarithmic corrections. 
 For no-slip boundary conditions, the exact solution derived by \citet{Wen22} shows a weaker $ Re $ scaling, with $ \beta = 0.47$ 
 for rolls with $ Nu $-maximising aspect ratios  which is close to the value $ \beta = 0.46 $ which was observed in  3-D DNS \citep{Iyer20} up to $ Ra = 10^{15}$.  
  
In a previous paper \citep{Lindborg23}, it was argued that the velocity and thermal boundary layer widths in the 3-D system are proportional to the smallest length scales that can develop in 3-D turbulence, that are the Kolmogorov and Batchelor scales. Using this assumption, it was shown that classical Nusselt number scaling is recovered in the limit of high $ Ra $.  Moreover, the fundamental scaling relation of 3-D turbulence was used to deduce the Reynolds number scaling. This relation states that
\begin{equation} \label{Fundamental}
\frac{\epsilon}{E^{3/2}/L} = {\mbox{Constant}}   \hskip 5mm  {\mbox{as}} \hskip 5mm  Re =  \frac{E^{1/2} L}{\nu} \rightarrow \infty  \, ,
\end{equation}
where $ \epsilon $ is mean kinetic energy dissipation, $ E $ is mean turbulent kinetic energy and $ L $ is the turbulent integral length scale. The relation (\ref{Fundamental}) is most often written using $ u^{3} $ instead of $ E^{3/2} $, where $ u $ is a characteristic turbulent velocity. It  has been experimentally verified  in a wide range of turbulent flows \citep{Sreeni98} and numerically verified in 3-D RBC  \citep{Pandey22}. Using (\ref{Fundamental}) and the equations of motions it is straightforward to derive $ Re \sim (Ra Nu Pr^{-2})^{1/3} $. Assuming $ Nu \sim Ra^{1/3} $ we thus have 
\begin{equation} \label{Reynolds3D}
Re \sim Pr^{-2/3} Ra^{4/9} \, ,
\end{equation} 
for the 3-D system. The relation (\ref{Reynolds3D}) was previously derived by a number of other investigators \citep[e.g.][]{Kraichnan62, Siggia94} under different assumptions. 
The scaling (\ref{Reynolds3D})  is in very good agreement with experimental and numerical data. 
 \citet{Ashkenazi99} report $ Re \sim Ra^{0.43}  $ from high Rayleigh number experiments and \citet{Iyer20} report $ Re \sim Ra^{0.46} $ from DNS of convection in a low aspect ratio cylindrical cell at $ Ra \in [10^{9}, 10^{15}] $. It is also interesting to note that the exact steady 3-D solution which recently was found by \citet{Motoki21} exhibits $ Ra^{4/9} $-scaling, while the 2-D solution derived by the same authors exhibits a considerably faster increase of $ Re $ with $ Ra $, unlike the solution derived by \citet{Wen22}. 
 
In this paper, we will argue that the Reynolds number scaling is different in 2-D compared to 3-D and that it is not consistent with the prediction of the theory of the ultimate state. 
 The reason for the difference in Reynolds number scaling between 2-D and 3-D is that 
dissipation is much weaker in 2-D than in 3-D and that (\ref{Fundamental}) cannot hold in 2-D.  
As will be argued, the weaker dissipation  also implies that the 2-D system will converge on a much longer time scale than the 3-D system. To deduce a lower bound on the time scale we will first derive an equation for the evolution of mean kinetic energy. In the end, based on a summary of observations from DNS, we will argue that classical Nusselt number scaling indeed holds in the limit of high $ Ra $, both in 2-D and 3-D, although the Reynolds number scaling and the convergence time scale are very different in the two systems.

\section{Evolution equation for the mean kinetic energy}
We assume that the flow is described by the Navier-Stokes equations under the Boussinesq approximation 
\begin{eqnarray}
\frac{{\mbox{D}} \bu}{{\mbox{D}} t}   =  - \frac{1}{\rho}  \nabla p + g \alpha T  {\bf e}_z +  \nu \nabla^2 \bu   \, , \\ \label{ND}
\nabla \cdot \bu  =  0 \, ,  \\ \label{Temp} 
\frac{{\mbox{D}} T }{{\mbox{D}} t} = \kappa \nabla^2 T\, ,
\end{eqnarray} 
where $ \rho, p, g, \nu $ and $ \kappa $ are density, pressure, acceleration due to gravity, kinematic viscosity and diffusivity, $ {\be }_z $ is the vertical unit vector,  $ T  $ is  temperature and  $ \alpha $ is the thermal expansion coefficient. 
We locate a lower boundary at $ z = -L/2 $, and an upper boundary at $ z = L/2 $, lateral boundaries at $ x = -X_0  $ and $ x = X_0  $. In three dimensions we also introduce lateral boundaries at $ y = -X_0  $ and $ y = X_0  $. We assume that  $ X_0  \gg L $,  so that horizontal mean values will be well converged.  We apply constant temperature boundary conditions at the lower and upper boundaries, with $ T = \Delta T/2 $ and $ T = -\Delta T/2 $ at the lower and upper boundaries, respectively. At the lateral boundaries we apply adiabatic boundary conditions. For the velocity field we apply stress-free or no-slip boundary conditions.  We assume that the initial temperature  is linear, $ T = -\Delta T z/L $, and that the initial velocity field is very close to zero.  The non-dimensional input parameters of the problem are the Rayleigh and Prandtl numbers, defined as
\begin{eqnarray}
Ra = \frac{g  \alpha \Delta T  L^3}{\nu \kappa}\, ,  \hskip 1cm   Pr = \frac{\nu}{\kappa} \, ,
\end{eqnarray} 
while the output parameters are the time-dependent Nusselt and Reynolds numbers defined as
\begin{equation}
Nu =  -{\frac{{\mbox{d}} \overline{T} }{{\mbox{d}} z} |_{z=-L/2} } / {( \Delta T /L )}  \, , \hskip 1cm Re = \frac{E^{1/2} L}{\nu}  \, , 
\end{equation} 
where $ E $ is the domain mean value of the kinetic energy per unit mass, and the bar denotes a horizontal mean.  
 
Using Cartesian tensor notation, the kinetic energy equation can be written as
\begin{equation} \label{Kinetic} 
\frac{1}{2} \pd{u_i u_i}{t}  = - \pd{}{x_j} \left (u_j \left (\frac{1}{2} u_i u_i + \frac{p}{\rho} \right ) \right ) + g \alpha T w - 2\nu S_{ij} S_{ij} + 2\nu \pd{}{x_j} \left (u_i S_{ij} \right ) \, , 
\end{equation} 
where $  w $ is the vertical velocity and $ S_{ij} $ is the strain rate tensor. In turbulence theory, dissipation is often expressed in terms of vorticity, $ {\boldsymbol \omega} $, rather than strain. For an incompressible fluid such a formulation is perfectly consistent, which can be seen from the identity
\begin{equation} \label{SV}
2\nu S_{ij} S_{ij} = \nu \omega_i \omega_ i +  2\nu \pd{}{x_j} \left ( u_i \pd{u_j}{x_i} \right ) \, .
\end{equation}
Since the last term will integrate to zero over a volume with no-slip or stress-free boundary conditions, dissipation can be defined as $ \nu \omega^2 $ instead of  $ 2 \nu S_{ij} S_{ij} $. 
This definition is preferable in the context of two-dimensional Rayleigh-B\'enard convection, for two reasons. First, conservation of enstrophy, $ \omega^2/2 $, is central in the theory of two-dimensional turbulence. To be able to link  dissipation to enstrophy has therefore 
certain theoretical advantages.  Second, linking dissipation to vorticity will clarify the difference between stress-free and no-slip boundary conditions. With stress-free conditions, vorticity  is zero at the boundaries in 2-D which is not generally true for $ S_{ij} S_{ij} $. 
A vorticity based definition will thus guarantee that boundary layer dissipation is small with stress-free conditions as opposed to the case with no-slip conditions. 

To derive the expression for the evolution of $ E $ we integrate the temperature equation (\ref{Temp}) to obtain
\begin{equation} \label{Help} 
\overline{wT} = - \int_{-L/2}^{z} \pd{\overline{T}}{t} \, {\mbox{d}}z  + \kappa Nu \frac{\Delta T}{L} + \kappa \pd{\overline{T}}{z} \, . 
\end{equation} 
Integrating (\ref{Kinetic}) over the whole domain, using (\ref{SV}) and (\ref{Help}), assuming 
that $ \overline{T} $  remains an odd function of $ z $ during the evolution of the flow and integrating in time with given initial conditions, we find 
\begin{equation} \label{Evolution} 
E(t) = \alpha g \int_{-L/2}^{L/2} \frac{z}{L} \left ( \overline{T}(z, t) + \frac{z}{L} \Delta T \right ) {\mbox{d}} z  + \int_0^t \left ( \frac{\kappa^2 \nu}{L^4} Ra (Nu(t)-1) - \epsilon(t) \right ) {\mbox{d}}t  \, ,
\end{equation} 
where $ \epsilon $ is the mean dissipation, which in two dimensions can be expressed as
\begin{equation}
\epsilon = \frac{1}{2LX_0} \int_{-X_0}^{X_0} \int_{-L/2}^{L/2} \nu  \omega^2  \, {\mbox{d}} z {\mbox{d}} x \, ,
\end{equation} 
with a corresponding expression in three dimensions. 
The first term on the right hand side of (\ref{Evolution}) arises from the first term on the right hand side of (\ref{Help}) in the following way
\begin{eqnarray}
\nonumber - \int_0^t  \frac{\alpha g}{L} \int_{-L/2}^{L/2}  \int_{-L/2}^{z} \pd{\overline{T}}{t} \, {\mbox{d}}z^\prime   {\mbox{d}}z{\mbox{d}} t  = \\  \nonumber - \frac{\alpha g}{L} \int_{-L/2}^{L/2}  \int_{-L/2}^{z} (\overline{T}(z^\prime, t)- \overline{T}(z^\prime, 0)) \, {\mbox{d}}z^\prime   {\mbox{d}}z    = \\
\nonumber 
-  \frac{\alpha g}{L} \left [ z  \int_{-L/2}^{z } (\overline{T}(z^\prime, t)- \overline{T}(z^\prime, 0)) \, {\mbox{d}}z^\prime  \right ]_{z  = -L/2}^{z  = L/2}  \\ \nonumber  + \frac{\alpha g}{L} \int_{-L/2}^{L/2} z  (\overline{T}(z, t)- \overline{T}(z, 0)) {\mbox{d}} z =  \\
\alpha g \int_{-L/2}^{L/2} \frac{z}{L} \left ( \overline{T}(z, t) + \frac{z}{L} \Delta T \right ) {\mbox{d}} z \, , 
\end{eqnarray} 
where it has been assumed that $ \overline{T}(z,t) $ is an odd function of $ z $ and that $ \overline{T}(z,0) = - \Delta T z/L $. 
Introducing the free-fall velocity, $ u_f = \sqrt{gL\alpha \Delta T} $, and the non-dimensional variables 
\begin{eqnarray}
\tilde{E} = \frac{E}{u_f^2} \, , \hskip 8mm  \tilde{\epsilon} = \frac{ L \epsilon}{u_f^3} \,, \hskip 8mm
\tilde{t} =  \frac{u_f t} {L}  \, ,  \hskip 8mm \tilde{\overline{T}} = \frac{\overline{T}}{\Delta T} \, , \hskip 8mm \tilde{z} = \frac{z}{L} \,  ,
\end{eqnarray} 
equation (\ref{Evolution})  can be written in non-dimensional form as 
\begin{equation} \label{Evolution2} 
\tilde{E}(\tilde{t}) =  \int_{-1/2}^{1/2} \tilde{z} ( \tilde{\overline{T}}(\tilde{z}, \tilde{t})+ \tilde{z}) \, {\mbox{d}} \tilde{z}  + \int_0^{\tilde{t}} [ Pr^{-1/2} Ra^{-1/2} (Nu(\tilde{t})-1) - \tilde{\epsilon}(\tilde{t}) ]  \, {\mbox{d}} \tilde{t} \, .
\end{equation} 
From (\ref{Evolution2}) it follows that in the stationary state we will have
\begin{equation}  \label{Production} 
\tilde{\epsilon} =  Pr^{-1/2} Ra^{-1/2} (Nu -1) \, , 
\end{equation} 
 which previously has been shown in many studies.

\section{Reynolds-number scaling and convergence time scale}
The key property distinguishing 2-D turbulence from 3-D turbulence is the conservation of enstrophy by the nonlinear term. The equation for mean enstrophy, $ \Omega = \langle \omega^2 \rangle/2 $, can be written as
\begin{equation} \label{Enstrophy} 
\pd{ \Omega }{t} = P-\epsilon_{\omega} + \frac{1}{2L  X_0} \int_{boundaries} \nu {\boldsymbol n} \cdot \nabla \left ( \frac{\omega^2}{2} \right )  {\mbox{d}} s \, , 
\end{equation} 
where $ P = -g \alpha  \langle \omega \partial_x T \rangle $ is mean enstrophy production by buoyancy, $ \epsilon_{\omega} = \nu \langle \partial_i \omega \partial_i \omega \rangle $ is mean enstrophy dissipation and the last term (where $ \boldsymbol n $ is the outwards pointing normal unit vector) is enstrophy production at the boundaries. 

With stress-free boundary conditions, the last term  is zero, because $ \omega $ is zero at the boundaries. For the same reason, kinetic energy dissipation, defined as $ \nu \omega^2 $, is zero at the boundaries and the contribution to total dissipation from the boundaries is negligible.  
As shown by \citet{Fjortoft53} and \citet{Kraichnan67} 
there can be no downscale energy cascade in a system where enstrophy is conserved by the non-linear term. Instead, energy will tend to cascade to larger scales. In 3-D RBC, the kinetic energy spectrum peaks at a wavenumber that is inversely proportional to $ L $ and falls off as $ \sim k^{-5/3} $ at higher wave numbers (e.g. Pandey 2022), indicating that energy is predominantly injected at scales that are proportional to $ L $. Most likely,  the energy injection scale  is proportional to $ L $ also  in the 2-D system, which leaves no room for an extended inverse energy cascade range. At large wave numbers, the energy spectrum of the 2-D system, most likely,  falls off as $ \sim k^{-3} $, as predicted by Kraichnan (1967), or somewhat steeper, as observed in most simulations of 2-D turbulence, especially simulations dominated by coherent structures. A surprisingly clean $ k^{-3} $-spectrum was observed in simulations of high Rayleigh number 2-D RBC performed by \citet{Samuel23}.  With a clean $ k^{-3} $-spectrum, mean enstrophy will scale as $ \Omega \sim E/L^2 \ln(L/\eta_\Omega) $ where $ \eta_\Omega = \nu^{1/2} \epsilon_{\Omega}^{-1/6} $, and $ \epsilon_{\Omega} $ is the enstrophy dissipation rate. With a steeper spectrum,  mean enstrophy will scale as $ \Omega \sim E/L^2 $, without the logarithmic factor. 
For the purpose of the present analysis, it is of limited importance whether the logarithmic factor is included or excluded, why we choose to exclude it. 
It can be concluded that if the energy injection scale is proportional to $ L $, 
energy as well as enstrophy will  accumulate in the lowest order modes.  Convection rolls that extend over the whole domain can be seen as a manifestation of this. 
This is the only assumption that is needed in order to derive the Reynolds number scaling. 
Kinetic energy dissipation then scales as
\begin{equation} \label{Dissipation}
\epsilon = 2 \nu \Omega \sim \nu \frac{E}{L^2} \hskip 5mm \Rightarrow \hskip 5mm \frac{\epsilon}{E^{3/2}/L} \sim Re^{-1} \  \, .
\end{equation} 
In the 2-D system, the  scaling of dissipation is thus strikingly different from the scaling (\ref{Fundamental}) in the 3-D system. Although the argument for (\ref{Dissipation}) seems very strong in the case with stress-free boundary conditions it can be questioned that (\ref{Dissipation}) also holds in the case with no-slip conditions. The production of enstrophy at the boundaries by the last term in (\ref{Enstrophy}) could give rise to a considerably larger dissipation in the boundary layers than in the central region. Assuming that the characteristic boundary layer dissipation is limited by the dissipation at the wall, we obtain  $ \epsilon_{bl} \lesssim  {u_{\tau}}^{4} \nu^{-1} $. With a friction law of the form $ u_{\tau} \sim U Re^{-\mu} $ and a boundary layer thickness scaling as  $ \delta \sim \nu {u_{\tau}}^{-1} $, an upper bound of the ratio between the total boundary layer dissipation and the dissipation in the interior can be estimated as $ R_{2D} \lesssim Re^{1-3\mu } $ in 2-D,  and $ R_{3D} \lesssim Re^{-3\mu}  $ in 3-D. Using a different line of reasoning involving the estimate $ \epsilon_{bl}  \sim \nu U^2/\delta^2 $, \citet{Lindborg23} arrived at the estimate $ R_{3D} \sim Re^{-1/4} $ in the 3-D case. As pointed out by one of the reviewers, the estimate for $ \epsilon_{bl} $ used by \citet{Lindborg23} may have been too naive.  In 2-D, the same estimate can definitely not be used, due to the stronger degree of inhomogeneity close to the wall in 2-D as compared to 3-D.  As pointed out in the introduction, experiments and numerical simulations show that $ \mu = 1/4 $ in 2-D Poiseuille flow. Arguably, the wall shear stress cannot be larger in a buoyancy dominated flow than in a shear dominated flow, which gives us the upper bound $ R_{2D} \lesssim Re^{1/4} $. 
 DNS data by  \citet[][figure 6]{Zhang17} show that $ Ra_{2D} $ is constant and approximately equal to $ 1.5 $ over four orders of magnitude in $ Ra $ ($ Ra \in [10^6, 10^{10}] $).  It seems unlikely that $ R_{2D} $ should increase dramatically for higher $ Ra $. 
 We therefore assume that $ R_{2D} $ is Reynolds number independent and that (\ref{Dissipation}) therefore also holds in the case with no-slip conditions. The data of  \citet[][figure 5 (a,b)]{Zhang17} show that dissipation is  approximately constant in the central region, with a sharp increase  at the edge of the boundary layers. Exact solutions with stress-free boundary conditions \citep{Chini09} also show that enstrophy is approximately constant in the central region but with a sharp decrease at the edge of the boundary layers. In both cases, it can be assumed that the thermal boundary layer width, $ \delta_T $,  is determined only by $ \kappa $ and $ \Omega $. 
 Dimensional considerations then give $ \delta_T \sim \kappa^{1/2} \Omega^{-1/4} \sim \kappa^{1/2} \nu^{1/4} \epsilon^{-1/4} $, which is the Batchelor scale \citep{Batchelor59} calculated using the mean dissipation of the system. As pointed out by \citet{Lindborg23} this is exactly the scale  of $ \delta_T $ that corresponds to  classical Nusselt number scaling.

Using (\ref{Production}), with $ Nu-1 \approx Nu $, (\ref{Dissipation}), and $ Nu \sim Ra^{1/3} $, we obtain
\begin{eqnarray}  \label{EN}
\tilde{E} & \sim & Pr^{-1} Nu \sim Pr^{-1} Ra^{1/3} \, ,  \\ \label{Reynolds} 
Re & \sim & Pr^{-1} Ra^{1/2} Nu^{1/2} \sim  Pr^{-1} Ra^{2/3} \, ,
\end{eqnarray} 
which are two equivalent expressions of the same identity. It may be of some interest to point out that if we had assumed the boundary layer dissipation is dominant in the no-slip case  and that $ R_{2D} $ reaches the upper limit $ Re^{1/4} $, we would  have obtained
$ Re \sim Ra^{16/27} Pr^{-8/9} $ instead of (\ref{Reynolds}) in the no-slip case. 
The Reynolds number scaling (\ref{Reynolds}) is identical to the expression derived for exact solutions with stress-free boundary conditions  by \citet{Wen20}. 
 If we instead had assumed that the Nusselt number scales in accordance with the theory of the ultimate state, 
the same line of reasoning had lead to an even stronger scaling, $ Re \sim Ra^{3/4} $, with some  Prandtl number dependent factor and a possible logarithmic factor. 
This is in contradiction with the Reynolds number scaling predicted by the same theory.

Substituting (\ref{EN}) into the left hand side of (\ref{Evolution2}) we can determine a minimum time it will take for the system to settle. 
An upper limit of the first term on the right hand side of (\ref{Evolution2}) can be estimated by putting $  \tilde{\overline{T}} = 0  $ in the integral,  which in this case will be equal to $ 1/12 $.  This estimate is not crucially dependent on the assumption that the initial temperature profile is linear. The closer the initial temperature profile is to the final stationary profile, the smaller is this term. 
The first term on the right hand side of (\ref{Evolution2}) is thus negligible in comparison to the left hand side which is of the order of $ Pr^{-1} Nu \gg 1 $.  Assuming that $ Nu(\tilde{t}) < C Nu_{st} $ where  $ C $ is a constant and $ Nu_{st} $ is the Nusselt number in the stationary state, we obtain 
\begin{equation} \label{Long}
\tilde{\tau}  > c Pr^{-1/2}Ra^{1/2} \, .
\end{equation}
where $ c $ is a non-dimensional constant.  Using plots of the evolution of mean kinetic energy from the simulations at $ Ra \leq 10^{11} $ of \citet{Zhu18} communicated to the author by Zhu \& Lohse, the constant can be estimated as $ c \approx 0.01 $.  To reach a stationary state in a simulation at $ Ra = 10^{14}, Pr =1  $, would thus require of the order of one hundred thousand nondimensional time units. The simulation by \citet{Zhu18} at $ Ra = 10^{14}, Pr = 1 $, was run for two hundred and fifty nondimensional time units. 
It is noteworthy that (\ref{Long}) is derived using only (\ref{Evolution2}), (\ref{Dissipation}) and the assumption $ Nu(\tilde{t}) < C Nu_{st} $,  but no assumption regarding the Nusselt number scaling. In dimensional form (\ref{Long}) is expressed as $ \tau > cL^2/\nu $,  which is a general expression for the convergence time scale of a 2-D system undergoing an inverse energy cascade. 
 
 For the 3-D system it is not as straightforward to estimate $ \tilde{\tau} $. Most likely, $ \tilde{\tau} $ increases with $ Ra $ also in the 3-D system (private communication with Jörg Schumacher), although not as fast as in the 2-D system. 
 In 3-D the left hand side of (\ref{Evolution}) is of the order of $ Ra^{-1/9} Pr^{-1/3} $. Formally, it is therefore a subleading term in the limit of high $ Ra $ since the first term on the right hand side is independent of $ Ra $ and therefore, formally,  is of the order unity, although it is smaller than $ 1/12 $. The Prandtl number dependence complicates the matter and we therefore only make an estimate for $ Pr = 1 $. In this case 
 stationarity can not be reached until the dissipation term is of the order of unity. Assuming that mean dissipation, $ \tilde{\epsilon} $, during the evolution of the flow is limited by the value it takes in the final state we obtain $ \tilde{\tau} \gtrsim Ra^{1/6} $ for the 3-D system. 

The total computational cost in a simulation is proportional to the number of grid points multiplied by the number of time steps. For $ Pr \sim 1 $, the temperature and velocity fields should be resolved at the Kolmogorov scale, which will require that the number of grid points in each direction is proportional to $ Ra^{1/3} $.   Assuming that $ \tilde{\tau} \sim Ra^{1/2} $ in 2-D and $ \tilde{\tau} \sim Ra^{1/6} $ in 3-D the total computational cost will  scale as
$ Ra^{7/6} \times N_{\Delta t} $ in both cases, where $ N_{\Delta t} $ is the number of time steps per non-dimensional unit time, that surely is as large in 2-D as in 3-D. As a matter of fact, it is likely that $ N_{\Delta T} $ will be larger in 2-D than in 3-D. A Courant condition based on the magnitude of the velocity will require a smaller time step in 2-D than in 3-D. The total cost will thus scale at least as fast with $ Ra $ in 2-D as in 3-D. 
At $ Ra = 10^{14} $, it may actually be more costly to run a fully resolved simulation to a stationary state in 2-D than in 3-D.

\section{Comparison with DNS data} 
We first consider the $ Nu $ and $ Re $ scaling for the case with stress-free boundary conditions and then move to the case with no-slip  conditions. In each case, we  consider i) $ \gamma $ in $ Nu \sim Ra^{\gamma} $, ii) whether $ Nu $ is independent of $ Pr $, iii) $ \beta $ in $ Re \sim Pr^{\sigma} Ra^{\beta} $ and iv) $ \sigma $ in the same expression. Finally, we consider evidence of the scaling of the convergence time scale.  

\subsection{Stress-free boundary conditions}
\citet{Wang20} made an extensive DNS study of 2-D RBC with stress-free boundary conditions and studied the flow evolution for different roll states, quantified by the roll aspect ratio, $ \Gamma_r \in [1.6, \, 8] $, with $ Ra $ and $ Pr $ systematically varied in the ranges $ Ra \in [10^{7}, \, 10^{9}] $ and $ Pr \in [1, \, 100] $. 
\begin{enumerate}
\item  The value of $ \gamma $ was slightly increasing with $ \Gamma_r $,  within the range $ \gamma \in  [0.302, 0.321] $. 
\item The value of $ Nu $ was virtually independent of $ Pr $. 
\item The value of $ \beta $ was slightly increasing with $ \Gamma_r $, within the range $ \beta \in [0.657, 0.675] $. 
\item The value of $ \sigma $ was slightly increasing with $ \Gamma_r $, within the range $ \sigma \in [-1.078, -1.043] $. 
\end{enumerate}
In conclusion, the data of \citet{Wang20} strongly suggest that the 2-D system approaches classical Nusselt number scaling and Reynolds number scaling (\ref{Reynolds}) at $ Ra \sim 10^{9} $,  as also pointed out by \citet{Wen20}.  
 
 \subsection{No-slip boundary conditions} 
  \begin{enumerate}
\item
 \citet{Johnston09} report $ \gamma = 0.285 $ at $ Pr = 1 $ and $ Ra \in [10^{7}, \, 10^{10}] $.  \citet{Zhang17} report $ \gamma = 0.3 $ for $ Pr =  0.7 $ and $  5.3 $ at $ Ra \in [10^{6}, \, 10^{10}] $. 
 \citet{Wang20b} plot $ Nu/Ra^{1/3} $ at $ Pr = 10 $ in a lin-log plot and find that $ \gamma $ is slightly increasing from $ \gamma = 0.262 $ in $ Ra \in [10^{7}, \, 10^{8}] $ to $ \gamma = 0.289 $ at $ Ra \in [10^9, \, 10^{10}] $. \citet{Poel13} plot $ Nu/Ra^{1/3} $ at $ Pr = 4.38 $ in a lin-log plot showing a slightly convex curve at $ Ra \in [10^{7}, \, 10^{10} ]$. If extrapolated to higher $ Ra $, the curve would approach a straight line at $  Ra \sim  10^{14} $.  \citet{Pandey25} undertook very long simulations  up to $ Ra = 10^{10} $ for $ Pr = 0.1 $ and up to $ Ra = 10^{12} $ for $ Pr = 1 $ and report $ \gamma = 0.27 \pm 0.007 $ and $ \gamma = 0.29 \pm 0.003 $ in the two cases, respectively. 
 \citet{Zhu18} claim that they have determined $ Nu $ up to $ Ra= 10^{14} $ and \citet{Zhu19} that they have even reached $ Ra = 4.64 \times 10^{14} $. They find that $ \gamma = 0.357 $ at $ Ra > 10^{13} $, which they interpret as evidence of a transition to the ultimate state. It is difficult to see that $ \gamma = 0.357 $  can be invoked as such evidence since the Nusselt number in the ultimate state should scale as $ Ra^{1/2} $, including a possible logarithmic correction. However, the interpretation is made using an idea developed by \citet{Grossmann11} according to which, thanks to a possible logarithmic correction, it would be sufficient to measure an `effective scaling exponent' of $ \gamma = 0.38 $ in order to verify the existence of the ultimate state. The line of argument seems to be the following: Since $ \gamma = 0.357 $ is so close to the `effective scaling exponent' $ \gamma = 0.38 $, the issue has been settled, although $ \gamma = 0.357 $  is still very close to $ \gamma = 1/3 $. 
 All the data of \citet{Zhu18} and \citet{Zhu19} for $ Ra > 10^{10} $ were evaluated in states that were very far from stationarity (private communication with Detlef Lohse and Xiaojue Zhu). 
 Lohse \& Zhu claim (private communication) that the Nusselt number is converged, although the mean kinetic energy is very far from having reached stationarity. As evidence they point out that  in a simulations at $ Ra = 10^{11}  $ that was ran to stationarity after publication of \citet{Zhu18}, $ Nu $ increased only by three percent during the last phase of the simulation. It is true that the plots communicated to the author by Zhu \& Lohse indicate that the Nusselt number is almost converged while there is still a rapid growth of the mean kinetic energy in the simulations at $ Ra \le 10^{11} $.  However, they also show that the Nusselt number in the simulation at $ Ra = 10^{11} $ was not converged at the non-dimensional time, $ \tilde{t} = 250 $, at which the simulation at $ Ra = 10^{14} $ was ended. 
 In the author's opinion, there is no other way to test whether $ Nu $  is converged to the accuracy that is needed in order to distinguish 
 $ \gamma = 1/3 $ from $ \gamma = 0.357 $ at $ Ra >  10^{13} $ than running the simulations at $ Ra > 10^{13} $ for a considerably longer time.  
 In recent 2-D simulations carried out by \citet{He24}, the Nusselt  number was calculated up to $ Ra= 10^{13} $. As seen in their figure 5b,  their Nusselt number curve
 falls on top of the curve of \cite{Zhu18} up to $ Ra = 10^{11} $, while it falls above at $ Ra \in [10^{11}, \, 10^{13} ] $, where it conforms to $ Nu \sim Ra^{1/3} $.  Most likely, the reason behind this difference is that the Nusselt number in the simulations by \citet{He24} was  better converged than in the simulations by \citet{Zhu18}.

\item
\citet{Poel13} present a plot showing that $ Nu $ is virtually independent of $ Pr $ at $ Ra = 10^{8} $ and $ Pr \in [0.3, \, 100] $. The two curves of $ Nu $ by \citet{Zhang17} for $ Pr = 0.7 $ and $  5.3 $ are indistinguishable.  

\item  
\citet{Zhang17} report $ \beta = 0.6 $ for $ Pr =  0.7 $ and $  5.3 $. 
 \citet{Wang20b} observed a slight increase of $ \beta $ at $ Ra \in [10^{7}, \, 10^{10} ] $ and $ Pr = 10 $ with $ \beta = 0.565 $ for $ Ra \in [10^{7}, \, 10^8] $ and $ \beta = 0.595$ for $ Ra \in [10^{9}, \, 10^{10}] $. 
 \citet{Pandey21} report $ \beta = 0.6 $ at $ Ra \in [10^{6}, \, 10^{9}] $ and $ Pr = 0.021 $. 
 Although these investigations were carried out with great care, it seems likely that somewhat higher values would have been obtained if the simulations had been carried out for a very long time. 
\citet{Pandey25} undertook very long simulations at $ Pr = 0.1 $ and $ Pr = 1 $ and carefully monitored the convergence. Figure show $ Re Pr Ra^{-2/3} $ versus $ Ra $, from their simulations, showing the same data as their figure 7 which is an uncompensated plot. 
 Indeed, 
the data are in quite good agreement with $ Re \sim Ra^{2/3} Pr^{-1} $. However, \citet{Pandey25} find that the best power law fit is $ Re \sim Ra^{0.65\pm 0.01} $ for both Prandtl numbers. 
That $ \beta $ is still somewhat smaller than $ 2/3 $ can be expected. From (\ref{Reynolds}) it is clear that $ \beta $ will not reach $ 2/3 $ until $ \gamma $ reaches $ 1/3 $. In the limit of high $ Ra $ we should have $ \beta = (1+\gamma)/2 $. With $ \gamma =  0.29 \; (Pr = 1) $ and $ \gamma = 0.27 \; (Pr=0.1) $, as obtained by \citet{Pandey25},  our analysis predicts 
$ \beta = 0.645 $ and $ \beta = 0.635 $ in the two cases, respectively. This  in good agreement with $ \beta = 0.65 $.

\item  
\citet{Poel13} present a log-log plot of $ Re/Pr^{3/4} $ at $ Pr \in [0.1, \, 60] $. From the slope of the curve it can be estimated that $ \sigma \approx -0.9 $, which is not too far from $-1 $.  From the Nusselt number plot given by \citet{Zhang17} at $ Pr = 0.7 $ and $ Pr = 5.3 $ it can also  be estimated that $ \sigma  $ is not very far from $ -1 $. In figure 1, we see that the the two curves for $ Pr = 1 $ and $ Pr = 0.1 $ from the data by \citet{Pandey25}, depicting $ Re Pr $ versus $ Ra $, collapse quite nicely, which is consistent with $ \sigma = -1 $. 
\end{enumerate}
 
 In conclusion, we find it likely but not certain that  $ Nu $ will approach  $ Ra^{1/3} $, independent of $ Pr $ and that $ Re $ will approach $ Ra^{2/3}Pr^{-1} $ in the limit of high $ Ra $, 
 also in the case with no-slip boundary conditions.  

\begin{figure}
\begin{center}
\includegraphics[angle=0,width= 9 cm]{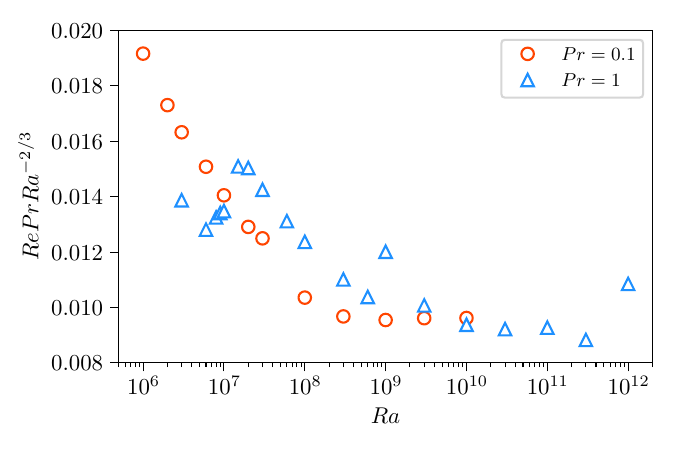}
\caption{ $ Re Pr Ra^{-2/3} $ versus $ Ra $ at $ Pr = 0.1 $ and $ Pr = 1 $ from DNS of \citet{Pandey25}. The figure was produced by  A. Pandey and K. Sreenivasan and is reproduced by their permission. }
\end{center}
\end{figure}

\subsection{Convergence time scale}
 The prediction (\ref{Long}) of the slow convergence of the mean kinetic energy is supported by a figure communicated to the author by Zhu \& Lohse, showing the time evolution of $ \tilde{E} $ from the four simulations (number 7,8,9 and 10) at $ Ra \in [10^{10}, \, 10^{11} ] $ and $ Pr =1 $ that were reported by \citet{Zhu18}.  The simulations were continued after publication to investigate the convergence of $ \tilde{E} $.  According to Zhu \& Lohse, the figure is `preliminary' and `not meant for publication'.
The prediction (\ref{Long})  is also supported by recent results by \citet{Pandey25} who carried out very long simulations up to $ Ra =  10^{10} $ at  $ Pr = 0.1 $ and up to $ Ra = 10^{12} $ at $ Pr = 1 $. They carefully estimated
$ \tilde{\tau} $ and found that $ \tilde{\tau} \sim Ra^{0.57} $ for $ Pr = 0.1 $ and $ \tilde{\tau} \sim Ra^{0.65} $ for $ Pr = 1 $, indicating that the approach to a stationary state at these $ Ra $ is even slower than the lower bound (\ref{Long}). They also found that $ \tilde{\tau} $ is larger for $ Pr =0.1 $ than for $ Pr = 1 $ at a fixed $ Ra $, which is in qualitative agreement with (\ref{Long}). 

 \section{Conclusions} We have shown that the Reynolds number cannot scale in the same way in the 2-D and 3-D RBC systems if the Nusselt number scales in the same way. 
  Assuming classical Nusselt number scaling in both cases, we obtain $ Re \sim Pr^{-1} Ra^{2/3} $ in 2-D compared to $ Re \sim Pr^{-2/3} Ra^{4/9} $ in 3-D.
 Results from 2-D DNS with  stress-free boundary conditions show very good agreement with $ Nu \sim Ra^{1/3} $ and $ Re \sim Pr^{-1}Ra^{2/3} $  while results from DNS with no-slip boundary conditions show that these scalings are almost reached at the simulated $ Ra $ also in this case.  The derivation given in this paper together with the DNS results reported by \citet{Pandey25}  show that  the Reynolds number scaling predicted by the ultimate state theory is far from valid in 2-D.  The theory is  therefore not applicable to 2-D RBC. 

Using the scaling $ \tilde{E}  \sim Pr^{-1} Nu $ and the equations of motion we deduced a lower bound of the convergence time scale, $  \tilde{\tau}  >  c Pr^{-1/2}Ra^{1/2} $, without making any assumption regarding the Nusselt number scaling.  The slow convergence is confirmed by results communicated to the author by Zhu \& Lohse and recent results by \cite{Pandey25}. 
From a computational point of view the slow convergence is, of course, disappointing. The general motivation for carrying out 2-D simulations is that they are supposed to reach the same $ Ra $ as 3-D simulations, but at a lower computational cost.   If this is not true, there is a risk that very few 2-D DNS will be carried out at $ Ra > 10^{10} $ in the future, which would be a pity. To overcome the convergence barrier it is necessary to develop smart strategies. One such strategy, used by \citet{Pandey25}, is to run the simulations at low resolution until a stationary state is reached after which the resolution is increased. 

The similarity of the scalings between the 2-D systems with stress-free and no-slip boundary conditions suggests  that wall shear stress is a relatively unimportant factor determining the dynamics. 
Most likely, this is also the case in 3-D.
According to the theory of the ultimate state,  as expounded by \citet{Lohse23}, the no-slip  system will undergo a shear induced  transition at some high Rayleigh number after which the heat flux will radically increase. Such a scenario seems unrealistic, given the similarity between the observed Nusselt and Reynolds number scalings in the no-slip and the stress-free systems.  Comparing the Nusselt number plots of  \citet[][figure 8c]{Wang20} with the plots of \citet[][figure 2]{Zhang17} it can be seen that the Nusselt number is actually  twenty to one hundred percent larger in the stress-free simulations than in the no-slip simulations at $ Ra \in [10^7, 10^{9}] $, suggesting that wall shear stress is reducing the heat transfer rather than reinforcing it.

The difference in Reynolds number scaling and convergence time scale  between the 2-D and 3-D systems is a reflection of the fundamental difference between the dynamics in the interior of a 2-D and 3-D convection cell.  The energy cascade goes in different directions in the two systems. In 2-D, enstrophy is conserved by the nonlinear terms and there is a downscale enstrophy cascade, whereas there is strong enstrophy production by the  nonlinear terms in 3-D. Yet, plots of $ Nu/Ra^{1/3} $ versus $ Ra $ are very similar in the two cases, with a slightly convex curve that seems to  approach a straight flat line in the limit of high $ Ra $. The  observation that the Nusselt number scaling is so similar in spite of the fact that the  interior dynamics is so different, supports the hypothesis \citep{Malkus54, Howard66} that the heat flux is not regulated by the interior dynamics but by convective instabilities in the boundary layers and that these instabilities are similar in the two cases. We should thus expect that the curve $ Nu/Ra^{1/3} $ will,  indeed,  approach a straight flat line in the limit of high $ Ra $, in accordance with the predictions of \citet{Malkus54} and \citet{Howard66}. The observation that the Nusselt number reaches an approximately constant value much faster than the mean kinetic energy in the 2-D system also supports the hypothesis that the heat transfer is mainly determined by processes in the boundary layers. As soon as they have developed, the heat transfer reaches an approximately constant value and is not seriously affected by the growth of the kinetic energy in the interior. 

In two thousand and twenty, Doering (2020) wrote an excellent short essay on the history of the two competing theories of heat transfer. Invoking the clean DNS results by \citet{Iyer20}, he concluded that `classical $ 1/3 $ scaling currently appears to be winning the competition'. 
 Since he wrote this, even more evidence supporting classical scaling has amounted.  \citet{Samuel25} showed that even in the case when there is an artificial shear mode included in a DNS of RBC, there is no sign of a transition to the ultimate state. 
\citet{Shevkar25} showed that thermal fluctuations in the boundary layers scale in excellent agreement with the assumption of marginal boundary layer stability underlying the prediction of classical scaling, and challenged the notion of a global boundary layer instability underlying the prediction of a transition to the ultimate state. \citet{Tiwari25a} performed DNS of compressible turbulence, and showed that classical scaling is approached also in this case. 
Just recently,  \citet{Tiwari25b}  made a comparison between the heat flux in compressible and standard Rayleigh-B\'enard convection observed in DNS, and developed a compelling general argumentation suggesting that the theory of the ultimate state is a blind alley. 

The observation that classical scaling  is approached also in the compressible case, suggests that it can be derived from general principles that do not rely on the Boussinesq approximation. A hint on what type of  principles we should look for is given by making the straightforward  observation that  $ Nu \sim Ra^{\gamma} $  implies that the dimensional heat flux scales as $ Q \propto \Delta T^{\gamma+1} L^{3 \gamma -1} $, if all parameters except $ L $ and $ \Delta T $ are unchanged.  If $ \gamma > 1/3 $,  a thinner convection cell would be more insulating than a thicker -- both filled with the same substance and exposed to the same boundary conditions. If $ \gamma > 1/3 $,  we would be able to carry out a series of experiments in which $ Q $ is kept constant, while $ L $ is {\em increased}  and $ \Delta T $ is {\em decreased} as $ \Delta T \propto L^{(1-3\gamma)/(\gamma+1)} $.   
In the author's mind, the possibility of such an experimental outcome is strongly counter intuitive, especially in the limit  of large $ L $, since the heat flux in this case would remain finite over an arbitrarily thick convection cell over which there is an infinitesimal temperature drop.  It remains a theoretical challenge to find the most general principles under which classical scaling can be derived. 
 \vskip 5mm
\noindent{\bf Acknowledgements} A. Pandey and K. Sreenivasan are gratefully acknowledged for sending me the plot shown in figure 1.   X. Zhu and D. Lohse are gratefully acknowledged for communicating information on their simulations. 
The author has benefited from discussions on RBC with K. Sreenivasan, J. Wettlaufer and J. Schumacher. 
Three anonymous reviewers of a previous version of this manuscript are acknowledged for constructive and useful criticism. Finally, the author would like to thank  F. Lundell for interesting discussions and thoughtful advice on research ethics. 
\vskip 2mm
\noindent{\bf Declaration of interest}  The author declares no conflict of interest.

\end{document}